\documentclass[prd, preprint, nofootinbib, showpacs]{revtex4}

\usepackage{graphicx}
\usepackage{amsfonts}
\usepackage{latexsym}
\usepackage{amssymb}
\usepackage{epsfig}

\begin{document}

\title{
 Originally asymmetric dark matter
}

\author{
Nobuchika Okada$^a$\footnote{okadan@ua.edu} 
 and Osamu Seto$^b$\footnote{seto@physics.umn.edu} 
  }

\affiliation{
$^a$Department of Physics and Astronomy, 
University of Alabama, Tuscaloosa, AL 35487, USA \\
$^b$Department of Life Science and Technology, 
Hokkai-Gakuen University, Sapporo 062-8605, Japan
}

%

\begin{abstract}

We propose a scenario with a fermion dark matter, 
 where the dark matter particle used to be the Dirac fermion, 
 but it takes the form of the Majorana fermion at a late time. 
The relic number density of the dark matter is determined 
 by the dark matter asymmetry generated through 
 the same mechanism as leptogenesis 
 when the dark matter was the Dirac fermion. 
After efficient dark matter annihilation processes 
 have frozen-out, a phase transition of a scalar field 
 takes place and generates Majorana mass terms 
 to turn the dark matter particle into the Majorana fermion. 
In order to address this scenario in detail, we propose two simple models. 
The first one is based on the Standard Model (SM) gauge group 
 and the dark matter originates the $SU(2)_L$ doublet Dirac fermion, 
 analogous to the Higgsino-like neutralino in supersymmetric models. 
We estimate the spin-independent/dependent elastic scattering 
 cross sections of this late-time Majorana dark matter 
 with a proton and find the possibility to discover it 
 by the direct and/or indirect dark matter search experiments 
 in the near future. 
The second model is based on the $B-L$ gauged extension of the SM, 
 where the dark matter is a SM singlet. 
Although this model is similar to the so-called Higgs portal dark matter 
 scenario, the spin-independent elastic scattering cross section 
 can be large enough to detect this dark matter 
 in future experiments.

\end{abstract}


\preprint{HGU-CAP 16} 

\vspace*{1cm}
\maketitle


\section{Introduction}

The origin of baryon asymmetry as well as of dark matter (DM) 
 are fundamental questions in cosmology.
Until now various mechanisms for the generation of baryon asymmetry and
 the candidates for dark matter have been investigated.
Under the standard paradigm, the dark matter is made up of
 the weakly interacting massive particle (WIMP) which freezes out from
 the thermal equilibrium, while the baryon asymmetry is supposed 
 to be generated by a mechanism that has nothing to do with 
 the relic abundance of the WIMP dark matter. 
However, the observed baryon density $\Omega_b$ and 
 that of dark matter $\Omega_{DM}$ are close to each 
 other~\cite{Komatsu:2010fb}, namely $\Omega_{DM}/\Omega_b \sim 5$, 
 and this fact might imply a certain connection between the two densities 
 in history of the Universe.  
For example, the $Q$-ball generated in the early Universe~\cite{Enqvist:1997si,Enqvist:1998en,Enqvist:2000gq,Roszkowski:2006kw,Seto:2007ym,Shoemaker:2009kg} 
 can be a single source to explain this concordance.

The so-called ``asymmetric dark matter" (ADM) scenario 
 offers another approach, where similarly to the relic baryon density, 
 the dark matter relic density is determined by a nonvanishing 
 asymmetry between dark matter particle and 
 its anti-particle~\cite{Barr:1990ca,Barr:1991qn,Kaplan:1991ah}, 
 rather than the usual thermal freeze-out via annihilation. 
In many ADM models proposed so far~\cite{
Thomas:1995ze, Hooper:2004dc, Kitano:2004sv, 
Agashe:2004bm, Cosme:2005sb, Farrar:2005zd, 
Khlopov:2005ew, Kaplan:2009ag}, the dark matter particle 
 shares a common quantum number with the baryon and 
 the dark matter asymmetry is generated associated 
 with the baryogenesis. 
For a successful ADM scenario, the following two conditions 
 should be satisfied. 
First, the dark matter particle is a complex field and 
 ``dark matter number'' can be defined which distinguishes 
 dark matter particle and its anti-particle. 
Second, the annihilation cross section between dark matter and 
 anti-dark matter particles is large enough for the symmetric 
 component of the abundance to be almost completely 
 erased~\cite{Cohen:2010kn,Buckley:2011kk,Graesser:2011wi,Iminniyaz:2011yp,MarchRussell:2012hi}. 
As a result, the relic abundance of dark matter is determined
 by the dark matter-antidark matter asymmetry left in the Universe. 
In some of ADM models, this condition of efficient annihilation 
 processes is achieved in the presence of 
 light mediator particles~\cite{Kaplan:2009ag,Ibe:2011hq}.

If an ADM has the electroweak interaction, especially 
 the coupling with Z boson with a large annihilation cross section, 
 the dark matter direct search experiments~\cite{HM} 
 give a very severe constraint on mass of the dark matter 
 $ \gtrsim {\cal O}$(10 TeV). 
This constraint can be avoided if a tiny dark matter 
 number violating mass term is introduced. 
However, it has been found that this term should be very small 
 $ \lesssim 10^{-41}$ GeV, in order to suppress 
 the dark matter-antidark matter oscillation in the early 
 Universe~\cite{Cohen:2009fz,Buckley:2011ye,Cirelli:2011ac,Tulin:2012re}.
While such a tiny dark matter number violation may be attractive 
 because the dark matter indirect detection becomes 
 available~\cite{Cohen:2009fz,Buckley:2011ye,Cirelli:2011ac}, 
 its smallness gives rise to the fine-tuning issue 
 in theoretical point of view.

In this paper, we propose a novel ADM scenario, 
 in which the dark matter relic density is 
 determined by the dark matter asymmetry in the early Universe, 
 but the nature of the particle is Majorana without introducing 
 such the extremely tiny parameter as mentioned above. 
The key ingredient is that the dark matter is the Dirac fermion 
 in the very early Universe, but a late-time phase transition 
 generates Majorana mass terms and as a result, 
 the initially Dirac fermion dark matter 
 takes the form of the Majorana fermion, 
 after the symmetric part of the DM abundance has been erased 
 by the annihilation between dark matters and anti-dark matters. 
In this paper we propose two simple models to address 
 this scenario in detail. 
In the first model (Model A) the dark matter particle has 
 the electroweak interaction through which the dark matter and  
 antidark matter particles annihilate very efficiently 
 in the early Universe. 
We consider the gauged $U(1)_{B-L}$ extension of the Standard Model (SM)
 for the second model (Model B), where the dark matter particle 
 is a singlet under the SM gauge group but has a $U(1)_{B-L}$ charge. 
For completeness, we also consider a concrete mechanism 
 for generating both the baryon and dark matter asymmetries. 
Various scenarios for baryogenesis by means of, for example,
 the out-of-equilibrium decay of heavy
 particles~\cite{An:2009vq,Haba:2010bm,Blennow:2010qp,Falkowski:2011xh},
 a scalar condensate~\cite{Bell:2011tn,Cheung:2011if,Graesser:2011vj,vonHarling:2012yn}
 and a CPT violating system~\cite{MarchRussell:2011fi,Kamada:2012ht} 
 have been proposed in the framework of the ADM scenario. 
Similarly to the scenario proposed in Ref.~\cite{Falkowski:2011xh}, 
 we adopt the same mechanism as leptogenesis~\cite{FukugitaYanagida}, 
 which is probably the simplest scenario for baryogenesis, 
 for generating the dark matter asymmetry as well as 
 the baryon asymmetry in the Universe.

The paper is organized as follows. 
We first consider Model A in the next section, 
 where new particles (Dirac fermions and complex scalars) 
 of the electroweak doublets and singlets are introduced. 
In the early Universe, the out-of-equilibrium decay of 
 the right-handed neutrinos generates the dark matter asymmetry  
 as well as the baryon asymmetry. 
The Dirac fermion dark matters efficiently annihilate 
 through the electroweak interaction and the resultant 
 relic abundance is determined by the dark matter asymmetry. 
A late-time phase transition by a light scalar generates 
 Majorana mass terms, by which the initially Dirac 
 fermion dark matter has taken the form of the Majorana particle. 
We estimate the elastic scattering cross sections of 
 the dark matter particle with a proton which are relevant 
 for the direct/indirect dark matter search experiments. 
In Sec.~III, we introduce Model B based on 
 the gauged $U(1)_{B-L}$ extension of the SM. 
New particles introduced in this model are all the SM gauge singlets. 
Our discussion on Model B is analogous to that on Model A, 
 but in Model B efficient annihilation processes of 
 Dirac fermion dark matters are realized by the resonance 
 effect of intermediate particles such as the $B-L$ gauge 
 boson and Higgs bosons. 
Although this model is similar to the so-called Higgs portal 
 dark matter scenario, we will see a crucial difference. 
Sec.~IV is devoted for summary and discussions. 
Our notations and the formulas used in our analysis 
 are listed in Appendices.

\section{Model A}
\subsection{Setup}

This model is based on the SM gauge group. 
We introduce a vectorlike pair of SU(2)L doublet fermions ($\psi_L$ and $\phi_R$), 
 a vectorlike pair of SM singlet fermions ($\chi_L$ and $\chi_R$), and
 SM singlet ($\phi$) and doublet complex scalars ($\eta$), 
 as well as the right-handed neutrinos ($N_R$). "
The charge assignments are shown in Table~I. 
Here, we have introduced a discrete $Z_2$ parity 
 to guarantee the stability of dark matter particle 
 and a global $U(1)$ symmetry to forbid Majorana mass terms
 for the gauge singlet fermions. 
The new doublet fermions are similar to the Higgsinos 
 in the minimal supersymmetric Standard Model (MSSM) 
 while the singlet fermions behave like a pair of Bino 
 without Majorana masses. 
The dark matter particle, which turns  out to be a mixture of 
 the doublet and singlet fermions after the electroweak symmetry 
 breaking, has a large annihilation cross section 
 through the electroweak gauge interaction. 
\begin{table}[ht]
\caption{Particle contents of Model A}
\centering
\begin{center}
\begin{tabular}{|l|c|c|c|c|c|} \hline
Fields  & $SU(2)_L$ & $U(1)_Y$ & $Z_{2}$ parity & Global U(1)   \\ \hline\hline
$\psi_L$ &   2       &  $+1/2$  &    $ - $          &    $-1$          \\ \hline
$\psi_R$ &   2       &  $+1/2$  &    $ - $          &    $-1$          \\ \hline
$\chi_L$ &   1       &     0       &    $ - $          &    $-1$          \\ \hline
$\chi_R$ &   1       &     0       &    $ - $          &    $-1$          \\ \hline\hline
$\Phi$    &   2       &  $+1/2$    &    $ + $          &    0               \\ \hline
$\eta$    &   2       &  $-1/2$     &    $ - $          &    $+1$          \\ \hline
$\phi$    &   1       &     0       &    $ + $          &    $+2$          \\ \hline\hline
$N_R$     &   1       &     0       &    $ + $          &    0               \\ \hline
\end{tabular}
\end{center}
\end{table}

The gauge invariant Lagrangian relevant to our discussion 
 is given by~\footnote{
To make our discussion simple, we have dropped 
  a term $Y_N^\prime \bar{N_R^c} \eta \psi_R$, 
 assuming $Y_N^\prime \ll Y_N$. 
}
\begin{eqnarray}
- {\cal L} & \supset &  \frac{1}{2}y \bar{\chi_R^c} \phi \chi_R + \frac{1}{2}y \bar{\chi_L^c} \phi \chi_L
 + \frac{1}{2} \bar{N_R^c} M_N N_R \nonumber \\
  & &  + \mu_{\chi} \bar{\chi_L} \chi_R + \mu_{\psi} \bar{\psi_L} \psi_R 
 + Y \bar{\psi_L} \Phi \chi_R + Y \bar{\chi_L} \Phi^\dagger \psi_R 
 + Y_N \bar{\psi_L} \eta^{\dagger} N_R 
 + {\it h.c.} \nonumber \\
&& + V(\Phi, \eta, \phi) 
\end{eqnarray}
 with
\begin{eqnarray}
V(\Phi, \eta, \phi) & = & m_1^2 |\Phi|^2 +\lambda_1 |\Phi|^4
       + m_3^2 |\phi|^2 + (\lambda_4 |\Phi|^2 + \lambda_6 |\eta|^2 ) |\phi|^2 
       + \lambda_7|\phi|^4 \nonumber \\
  && + m_4^2 |\eta|^2  +  \lambda_8 |\Phi|^2  |\eta|^2 
     + \lambda_{10}|\eta|^4  + {\rm H.c.}, 
\end{eqnarray}
where $\Phi$ denotes the SM Higgs doublet field. 
For simplicity, we have assumed common Yukawa coupling constants 
 ($y, Y$ and $Y_N$) between $\phi$ and $\chi_{L, R}$,
 between $\Phi$, $\chi_{L, R}$ and $\psi_{R,L}$, 
 and between $\eta$, $\psi_{L, R}$ and $N_R$, respectively. 
$M_N$ is the Majorana mass term of the right-handed neutrinos $N_R$.
We have omitted the flavor index for the right-handed neutrinos.

We adjust the parameters in the scalar potential 
 so as for $\Phi$ and $\phi$ (but not $\eta$) 
 to develop vacuum expectation values (VEVs) at a true vacuum, 
 where we expand the Higgs fields as 
\begin{eqnarray}
 \Phi = \left( 
            \begin{array}{c}
            0 \\
            \frac{1}{\sqrt{2}}(v + h) 
            \end{array}
              \right) , ~~\phi= \frac{1}{\sqrt{2}}(v_s + \varphi) ,
\end{eqnarray}
 with VEVs, $v= 246$ GeV and $v_s$.
The mass eigenstates are obtained by the unitary rotation 
\begin{eqnarray}
 \left( 
            \begin{array}{c}
            H_1 \\
            H_2 
            \end{array}
 \right)
   &=&
 \left( 
            \begin{array}{cc}
            S_{11} & S_{12}   \\
            S_{21} & S_{22}     
            \end{array}
 \right) 
   \left( 
            \begin{array}{c}
            \varphi \\
            h 
            \end{array}
  \right) 
 =
 \left( 
            \begin{array}{cc}
            \cos\alpha & \sin\alpha   \\
            -\sin\alpha & \cos\alpha     
            \end{array}
 \right) 
   \left( 
            \begin{array}{c}
            \varphi \\
            h 
            \end{array}
  \right)  ,
\end{eqnarray}
 with $\alpha$ being the mixing angle, which diagonalizes 
 the mass matrix 
\begin{equation}
 \textrm{ Mass term } = 
  \frac{1}{2} \left( \varphi \,\, h  \right)
  \left(
\begin{array}{ccc}
 2 v_s^2 \lambda_7   & v v_s \lambda_4 \\
 v v_s \lambda_4 & 2 v^2 \lambda_1
\end{array}
\right)
   \left( 
            \begin{array}{c}
            \varphi \\
            h
            \end{array}
  \right) . 
\label{HMassMatrix}
\end{equation}

In the following, we consider the phase transitions 
 by VEVs of $\Phi$ and $\phi$
 which take place in the early Universe 
 as the temperature of the Universe is decreasing 
 according to the expansion. 
Furthermore, we concentrate on a parameter space 
 where the electroweak symmetry breaking occurs 
 much earlier than the development of nonzero $\phi$ VEV.

\subsection{Cosmological evolution}
The right-handed neutrinos $N_R$ have two decay modes:
 decay into lepton and Higgs ($N_R \rightarrow L \Phi $), 
 and decay into $\eta$ and $\psi$ ($N_R \rightarrow \psi \eta $). 
Lepton asymmetry can be generated by the former decay mode 
 as in usual thermal leptogenesis \cite{FukugitaYanagida}.
The resultant asymmetry is expressed as
\begin{equation}
 \frac{n_L}{s} = \frac{\kappa_L}{g_*} \epsilon_{(N_R \rightarrow L \Phi)} 
 \times {\rm Br}(N_R \rightarrow L \Phi)
\end{equation}
 by using the dilution factor $\kappa_L$ 
 corresponding to the washout effect for 
 the generated lepton asymmetry, 
 the branching ratio of $N_R$ decay into lepton 
 ${\rm Br}(N_R \rightarrow L \Phi)$, 
 and the CP asymmetry parameter in the decay
\begin{equation}
 \epsilon_{(N_R \rightarrow L \Phi)} = \frac{\Gamma(N_R \rightarrow L \Phi)-\Gamma(N_R \rightarrow \bar{L} \Phi^* )}{\Gamma(N_R \rightarrow L \Phi)+\Gamma(N_R \rightarrow \bar{L} \Phi^*)} .
\end{equation}
%
The same mechanism generates the asymmetry between $\psi$ and $\bar{\psi}$ 
 such as 
\begin{equation}
 \frac{n_\psi}{s}
 = \frac{\kappa_{\psi}}{g_*} \epsilon_{(N_R \rightarrow \psi \eta)} 
 \times {\rm Br}(N_R \rightarrow \psi \eta). 
\end{equation}
Here $\kappa_{\psi}$ is the corresponding dilution factor, 
 the branching ratio of $N_R$ decay into $\psi$ 
 is ${\rm Br}(N_R \rightarrow \psi \eta )$, and
 the CP asymmetry in the decay is defined as 
\begin{equation}
 \epsilon_{(N_R \rightarrow \psi \eta )} = \frac{\Gamma(N_R \rightarrow \psi \eta)-\Gamma(N_R \rightarrow \bar{\psi} \eta^* )}{\Gamma(N_R \rightarrow \psi \eta)+\Gamma(N_R \rightarrow \bar{\psi} \eta^*)} .
\end{equation}
In the following, the dark matter particle is the Dirac fermion 
 which is a mixture of $\psi$ and $\chi$ 
 and therefore, this $\psi$-asymmetry generated by $N_R$ decay 
 is nothing but the asymmetry of $\psi$ dark matter sector. 
Notice that the asymmetry between $\eta$ and $\eta^*$ 
 is also generated.

After the electroweak symmetry breaking, 
 where $\phi$ has not developed its VEV yet,
 the mass matrix for the $Z_2$-odd Dirac fermions is given by
\begin{equation}
 (\bar{\chi_L}  \bar{\chi_R^c}  \bar{\psi_L}  \bar{\eta_R^c})
 M
\left(
\begin{array}{c}
 \chi_L^c \\
 \chi_R \\
 \psi_L^c  \\
 \psi_R \\
\end{array}
\right) ,
\end{equation}
 where
\begin{equation}
 M=
\left(
\begin{array}{cccc}
 0 & \mu_{\chi}  & 0 & Y \frac{v}{\sqrt{2}}  \\
 \mu_{\chi} & 0 & Y \frac{v}{\sqrt{2}} & 0   \\
 0 & Y \frac{v}{\sqrt{2}} & 0 & \mu_{\psi}   \\
 Y \frac{v}{\sqrt{2}} & 0 & \mu_{\psi} & 0   \\
\end{array}
\right) ,
\label{M-Dirac}
\end{equation}
 so that the mass eigenvalues are found to be 
\begin{eqnarray} 
 M^{Dirac}_{\chi\psi} = \frac{\mu_{\chi}+\mu_{\psi} \pm \sqrt{ 2 Y^2 v^2 + ( \mu_{\chi} - \mu_{\psi})^2 } }{2} .
\end{eqnarray}
Provided $\eta$ is heavier than those fermions, 
 the lightest mass eigenstate among the Dirac fermions 
 is the dark matter particle. 
Since this Dirac fermion partly consists of 
 the $SU(2)$ doublet fermions ($\psi$), its annihilation processes 
 are efficient in particular when the annihilation channels 
 into $WW$ and/or $ZZ$ are kinematically allowed.
This situation is analogous to the Higgsino-like neutralino 
 dark matter in the MSSM. 
Therefore, Dirac fermion dark matters and anti-Dirac fermion dark matters
 can almost completely annihilate and only the dark matter asymmetry is left 
 in the Universe. 
Similarly, the same is true for $\eta$ and $\eta^*$, hence
 its symmetric component of $\eta$ and $\eta^*$ is erased
 by annihilation through $SU(2)$ gauge interactions.

Since the $\phi$ is light, the scalar potential 
 at zero temperature is simplified at low energies 
 where the other scalars (and fermions) are decoupled: 
\begin{eqnarray}
V(\Phi, \eta, \phi) & \sim &
  - m_\phi^2 |\phi|^2 + \lambda_7|\phi|^4 .  
\label{A:LowPotential}
\end{eqnarray}
Here we have fixed parameters $ m_\phi^2 =- m_3^2- \lambda_4 v^2/2 > 0$ 
 and hence $\phi$ develops the VEV, 
\begin{equation}
 \langle \phi \rangle = \frac{v_s}{\sqrt{2}}
  = \sqrt{\frac{m_\phi^2}{2 \lambda_7}}. 
\label{A:VEV:phi}
\end{equation}
In the early Universe, this scalar potential is modified 
 by the thermal effect and temperature corrections are 
 given by~\cite{ThermalQFT} 
\begin{eqnarray} 
\delta V = 
 \frac{1}{64\pi^2} m^4 \left( \ln \frac{m^2}{\mu^2} -\frac{3}{2}\right) 
 + \frac{T^4}{2\pi^2} J_B(m^2/T^2) ,   
\end{eqnarray}
where $m^2= - m_\phi^2 + 4 \lambda_7 |\phi|^2$. 
In a high temperature, the function $J_B (m^2/T^2)$ 
 is expanded as 
\begin{eqnarray}
 J_B (m^2/T^2)=
 -\frac{\pi^4}{45} +\frac{\pi^2}{12}\left(\frac{m^2}{T^2}\right)
 -\frac{\pi}{6}\left(\frac{m^2}{T^2}\right)^{3/2}
 -\frac{m^4}{32 T^4}\ln\left(\frac{m^2}{a_B T^2}\right) + \cdots ,
\end{eqnarray}
where $\log a_B=5.4076$.
Since the scalar $\phi$ obtains the thermal mass 
 $m_{th}^2 \simeq T^2 \lambda_7/6$, 
 the finite temperature effective potential has a minimum at origin 
 when the temperature of the Universe is high. 
It is easy to find that $\phi$ becomes tachyonic below the temperature 
\begin{equation}
T_c \simeq \sqrt{\frac{6 m_\phi^2}{\lambda_7}} ,
\end{equation}
 and starts developing nonzero VEV. 
In this way, the Majorana masses of $\chi_{L,R}$ 
 are generated after this phase transitions 
 at a late time.

In this paper, we consider the case that 
 $T_c < M_{\chi\psi}^{Dirac}/20$, 
 which means the Dirac asymmetric dark matter particle 
 becomes Majorana at a late time well after the decoupling of the efficient 
 annihilation processes of the dark matters and antidark matters. 
The DM mass matrix is now modified to include the Majorana mass terms 
 such as 
\begin{equation}
 M=
\left(
\begin{array}{cccc}
 y \frac{v_s}{\sqrt{2}}  & \mu_{\chi}  & 0 & Y \frac{v}{\sqrt{2}}  \\
 \mu_{\chi} & y \frac{v_s}{\sqrt{2}} & Y \frac{v}{\sqrt{2}} & 0   \\
 0 & Y \frac{v}{\sqrt{2}} & 0 & \mu_{\psi}   \\
 Y \frac{v}{\sqrt{2}} & 0 & \mu_{\psi} & 0   \\
\end{array}
\right). 
\end{equation}
Therefore, the Dirac fermion asymmetric dark matter before the phase transition
 by $\langle \phi \rangle$ has taken the form of 
 the lightest Majorana fermion mass eigenstate 
 which is a linear combination among $\chi_{L,R}$ and $\psi_{L,R}$: 
\begin{equation}
 \chi_j = N_{j1} \chi_L^c + N_{j2} \chi_R + N_{j3} \psi_L^c +N_{j4} \psi_R , 
\end{equation}
 where $N_{ij}$ is the unitary matrix to diagonalize the mass matrix. 
When we take $\mu_{\psi}=\mu_{\chi} (=\mu)$ for example,  
 the mass eigenvalues are given by 
\begin{eqnarray} 
  \frac{ \sqrt{8 Y^2 v^2+ 2 y^2 v_s^2 } + \sqrt{2} y v_s}{4} \pm 4 \mu ,
\label{Simplified:Mchi}
\end{eqnarray}
%
%
%
and the lighter one is the dark matter mass.

The doublet scalar $\eta$ is heavier than the dark matter fermion 
 and decays into the dark matter and the SM particles. 
The main decay mode is found to be $\eta \to \chi_j {\bar \nu}$ 
 through the mixing between the heavy right-handed neutrinos 
 and the SM neutrinos, $\sqrt{\frac{m_\nu}{M_N}}$, 
 in the seesaw mechanism~\cite{seesaw}, 
 where $m_\nu$ is the typical light neutrino  mass scale. 
We estimate the decay width as 
\begin{eqnarray}
\Gamma(\eta\rightarrow {\bar \nu} \chi_j) 
 \simeq \frac{1}{16\pi} Y_N^2 M_{\eta}\left(\frac{m_{\nu}}{M_N}\right),
\end{eqnarray}
which corresponds to 
\begin{eqnarray}
\tau(\eta\rightarrow {\bar \nu} \chi_j) 
 \simeq 3.3 \times 10^{-5} \left(\frac{10^{-4}}{Y_N}\right)^2
 \left(\frac{0.1 {\rm eV}}{m_{\nu}}\right) \left(\frac{M_N}{M_{\eta}}\right) {\rm [sec]}.
\end{eqnarray}
We can choose the parameters so as for the $\eta$ 
 to decay after the decoupling of pair annihilation processes 
 of dark matters and also $\eta$s. 
As we have discussed  before, the $N_R$ decay generates 
 the same amounts for the dark matter asymmetry 
 and the $\eta$ asymmetry in the Universe. 
After the very efficient annihilation, 
 only the dark matters are left in the Universe, 
 in the same way, only the same amount of $\eta$s exist. 
The late-time $\eta$ decay creates the antidark matter 
 and, hence, the total dark matter number in the Universe 
 becomes zero after the $\eta$ decay. 
However, at that time, the dark matter pair annihilation 
 processes have already frozen-out, and 
 the total number density of the dark matter (plus anti-dark matter) 
 remains twice the number density of the dark 
 matter just before the decay of the $\eta$. 
%
%
%

Associated with the spontaneous global $U(1)$ symmetry breaking 
 by $\langle \phi \rangle \neq 0$, cosmic strings would be formed. 
However, this is cosmologically harmless 
 since the VEV scale of $\phi$ is small and 
 the mass per unit length of cosmic strings is very small. 
In addition, associated with this phase transition
 the NG mode appears.
However, this NG mode does not couple with SM particles,
 and the constraints from e.g., the cooling rate of stars 
 through the Nambu-Goldstone (NG) mode emissions inside the core 
 can be avoided. 
Since the NG mode was in thermal equilibrium, 
 its energy density contributes to the energy density of 
 the relativistic particles. 
We estimate the contribution to the extra neutrino species 
 at the big bang nucleosynthesis era $T_{BBN} \simeq 1$ MeV as 
\begin{eqnarray}
 \Delta N_{\rm eff} \simeq 
 \left.\frac{\rho_{\rm NG}}{\rho_{\nu}}\right|_{T_{BBN}} 
 \approx 0.2.
\end{eqnarray} 
This extra contribution is interesting in terms of 
 the value $N_{\rm eff} = 3.85 \pm 0.62$ 
 recently reported in Ref.~\cite{Keisler:2011aw} 
 using the data from the South Pole Telescope.

\subsection{Direct/Indirect detection of dark matter }
There is a variety of ongoing and planned experiments
 to detect a dark matter particle directly or indirectly, 
 through the elastic scattering of dark matter particle off 
 with nuclei. 
The dark matter in our model couples with quarks in two ways. 
One is through Higgs bosons relevant to the spin-independent 
 elastic scattering process, the other is the $Z$ boson exchange  
 which causes the spin-dependent elastic scattering 
 of the dark matter particle. 
Note that because of its Majorana nature, the dark matter 
 particle has only the axial vector coupling with the $Z$ boson, 
 and the $Z$ boson exchange process has no contribution 
 to the spin-independent elastic scattering process. 
If the dark matter is the Dirac fermion in the absence of $\phi$ VEV, 
 the $Z$ boson exchange dominantly contributes to 
 the spin-independent elastic scattering process 
 and as a result, our scenario will be excluded  
 as the dark matter scenario with the heavy Dirac neutrino DM~\cite{HM}
 or the left-handed sneutrino DM~\cite{Falk:1994es}~\footnote{
  Mixed~\cite{Hooper:2004dc,ArkaniHamed:2000bq} or 
  some right-handed~\cite{Lee:2007mt,Cerdeno:2008ep} sneutrinos 
  are still viable WIMP DM candidates.}.
Therefore, the late-time Majorana mass generation is crucial 
 for our dark matter scenario to be phenomenologically viable.

The cross section of the spin-independent elastic scattering 
 with a proton is given by 
\begin{equation}
\sigma_{\rm SI}^{(p)} = \frac{4}{\pi}
 \left(\frac{m_p m_{\chi}}{m_p + m_{\chi}}\right)^2 f_p^2, 
 \label{sigmaSI}
\end{equation}
 with the hadronic matrix element
\begin{equation}
 \frac{f_p}{m_p} = \sum_{q=u,d,s}f_{Tq}^{(p)}\frac{\alpha_q}{m_q} 
  + \frac{2}{27}f_{TG}^{(p)}\sum_{c,b,t}\frac{\alpha_q}{m_q},
\end{equation}
 where $m_q$ is a mass of a quark, 
 and $f_{Tq}^{(p)}$ and $f_{TG}^{(p)}$ are constants. 
$\alpha_q$ is the coefficient of the effective operator 
 ${\cal L}_{int} = \alpha_q \chi_j \chi_j \bar{q} q$ 
 for this scattering process and is given by 
\begin{eqnarray}
\alpha_q = \frac{1}{2}\sum_i 
  \frac{ (N_{j1}^2+N_{j2}^2) y S_{i1} +2(N_{j2}N_{j3}+ N_{j1}N_{j4}) 
  Y S_{i2}}{\sqrt{2} v M_{H_i}^2 } m_q S_{i2} . 
\end{eqnarray}
For $y v_s \ll Y v$, $\alpha_q$ is simplified as 
\begin{eqnarray}
\alpha_q & \simeq & \sum_i \frac{ y S_{i1} \pm Y S_{i2}}{4\sqrt{2} v M_{H_i}^2 }m_q S_{i2} \nonumber \\
 &=& \frac{m_q}{4\sqrt{2} v}\left( \frac{ y \cos\alpha \pm Y \sin\alpha}{ M_{H_1}^2 } \sin\alpha +  \frac{ - y \sin\alpha \pm Y \cos\alpha}{ M_{H_2}^2 } \cos\alpha \right) ,
\label{SimplifiedAlpha_q}
\end{eqnarray}
 where $\pm$ corresponds to which mass eigenvalue 
 in Eq.~(\ref{Simplified:Mchi}) is the dark matter mass.

When the Yukawa couplings are nonuniversal, 
\begin{eqnarray}
 {\cal L}{}_{Yukawa} & \supset & 
 Y_1 \bar{\psi_L} \Phi \chi_R + Y_2 \bar{\chi_L} \Phi \psi_R
 + {\it h.c.} , 
\end{eqnarray}
 we also consider the spin-dependent scattering processes of the dark matter. 
The corresponding cross section with $N$-nucleus 
 of mass $m_N$ and the spin $J$ is given by 
\begin{equation}
 \sigma_{\rm SD} = \frac{32}{\pi}G_F^2 
 \left(\frac{m m_N}{m+m_N}\right)^2 \Lambda^2 J (J+1) ,
\end{equation}
 with
\begin{equation}
 \Lambda = \frac{1}{J}( a_{\rm p} \langle S_{\rm p} \rangle + a_{\rm n}\langle S_{\rm n}\rangle ) ,
\end{equation}
 and the effective coupling with a proton and a neutron
\begin{eqnarray}
 a_{\rm p} &=& \sum_{q=u,d,s} \frac{\alpha_2}{\sqrt{2}G_F}\Delta_q^{(p)} , \\
 a_{\rm n} &=& \sum_{q=u,d,s} \frac{\alpha_2}{\sqrt{2}G_F}\Delta_q^{(n)} ,
\end{eqnarray}
 where $G_F$ is the Fermi constant, $\langle S_{\rm p (n)} \rangle$ 
 is the average of spin of proton (neutron) in a nuclei, 
 and $\Delta_q$ denotes the quark spin content.
$\alpha_2$ is the coefficient of the effective operator
 ${\cal L}_{int} = \alpha_2 \chi_j \gamma_5 \chi_j \bar{q} \gamma_5 q $ 
 mediated by the $Z$ boson, and this is proportional to $N_{j3}^2- N_{j4}^2$.
For $y v_s \ll Y v$, this is simplified as
\begin{equation}
 N_{j3}^2- N_{j4}^2 \simeq \frac{(Y_1^2 - Y_2^2)}
 { 2 \sqrt{(Y_1^2 - Y_2^2)^2 +8 (Y_1+Y_2)^2(\mu/v)^2 }}  .
\end{equation}
%
%
%

\subsection{Benchmarks}
There are a lot of parameters in our model 
 and we here choose typical parameter sets 
 for our benchmark scenarios. 
In order to realize our scenario, very efficient annihilation 
 processes of the asymmetric dark matters are necessary. 
In addition, it is interesting if there exists a parameter regions 
 which leads to the spin-independent/dependent elastic scattering 
 cross sections testable by the future experiments. 

\subsubsection{$M^{Dirac}_{\chi \psi} > M_W$ and $\alpha \simeq 0$}
\label{A:HeavyCase}

Firstly, we consider the case 
 where $M^{Dirac}_{\chi \psi} > M_W$ and
 the mixing angle of Higgs bosons is negligible, 
 $\alpha \simeq 0$~\footnote{
Since we have fixed $v_s \ll v$, 
 $\alpha \ll 1$ is natural for $\lambda_1 \sim \lambda_4 \sim
 \lambda_7$ from Eq.~(\ref{HMassMatrix}).
 }.
In this case, the Dirac fermion dark matters 
 (and also the $\eta$ scalars) sufficiently annihilate 
 into mainly $W^+ W^-$ through the $SU(2)$ gauge interaction 
 in the early Universe.
After the phase transition by nonzero $\langle \phi \rangle$, 
 the dark matter becomes Majorana but its relic number density 
 has been already determined by the dark matter asymmetry
 at the period when that particle was Dirac.
At present, since this dark matter particle is Majorana, 
 the cross section with nuclei which is relevant 
 to the dark matter direct detection is induced 
 by not the $Z$ boson exchange but Higgs boson exchanges.
For $\alpha\simeq 0$, the SM-like Higgs ($H_2$) exchange 
 process is dominant and we can use the simplified form 
 of $\alpha_q$ as 
\begin{eqnarray}
\alpha_q \simeq \frac{m_q}{4\sqrt{2} v} \frac{Y}{ M_{H_2}^2 } .
\end{eqnarray}
For instance, for $M_{H_2} \simeq 120$ GeV and $Y \simeq 0.2$,  
 we find 
\begin{equation}
\sigma_{\rm SI}^{(p)} \simeq 10^{-9} \rm{pb},
\end{equation}
 which is close to the current XENON 100 exclusion limit 
 of $\sigma_{\rm SI}^{(p)} \gtrsim 10^{-8}$ pb~\cite{Aprile:2011hi}.

Similarly, we estimate the spin-dependent cross section with a proton.
For instance, for $\mu \simeq 100$ GeV and $(Y_1, Y_2) \simeq (0.4, 0.2)$,
 we obtain
\begin{equation}
\sigma_{\rm SD}^{(p)} \simeq 3 \times 10^{-4} \rm{pb} ,
\end{equation}
 almost independently of the dark matter mass. 
The terrestrial experiments of dark matter direct search
 give the upper bound on $\sigma_{\rm SD}^{(p)} <$ several $\times 10^{-2}$ pb 
 by the SIMPLE~\cite{Girard:2011xc}. 
In fact, the most stringent upper bound is derived 
 from the indirect dark matter search by IceCube~\cite{Abbasi:2011zzr} as
%
 $\sigma_{\rm SD}^{(p)} \lesssim 10^{-3} - 10^{-4} \rm{pb}$, 
%
 depending on dark matter mass.
Our result can be close to this IceCube bound.

\subsubsection{$M^{Dirac}_{\chi \psi} < M_W$ and $\alpha\simeq 0$}
\label{A:LightCase}

Next, we consider the case where $M^{Dirac}_{\chi \psi} < M_W$ 
 and the mixing angle $\alpha \simeq 0$.
Since the annihilation channel to $W^+ W^-$ is kinematically 
 forbidden in this case, the Dirac fermion dark matters 
 mainly annihilate to a light $\phi$ pair through the Yukawa coupling. 
Since we have assumed the mass spectrum $M_{\eta} > M^{Dirac}_{\chi \psi}$,
 even in this case, we may expect the annihilation of $\eta$ into W-boson pairs.
The annihilation cross section of Dirac fermion dark matters is estimated as 
\begin{equation}
\sigma v \simeq \frac{9 (NyN)^4}{ 16 \pi m_{\chi}^2} \frac{T}{m_{\chi}} ,
\end{equation}
 which can be large enough for $y=0.1-1$. 
The spin-independent/dependent cross sections can be 
 the same as those in the case with $M^{Dirac}_{\chi \psi} > M_W$. 

\subsubsection{Sizable $\alpha$}
Finally we consider the case with a sizable $\alpha$.
In this case, the cross section with a proton can be enhanced
 by the first term in Eq.~(\ref{SimplifiedAlpha_q}) 
 because the singletlike Higgs ($H_1$) is light.
A portion of parameter space with very light $H_1$ 
 and/or not small $y$ can be excluded 
 by the current bound from the dark matter direct 
 search experiments.

\subsection{Precision measurement}
In our scenario, the dark matter particle is a mixture of 
 the $SU(2)$ doublet and singlet fermions 
 after the electroweak symmetry breaking, 
 which causes the isospin violation, 
 in other words, the mass splitting between  
 up-sector and down-sector fermions in the $SU(2)$ doublet. 
Such a mass splitting induces additional contribution  
 to the $\rho$ parameter, whose deviation from 1 is very severely 
 constrained, $\Delta \rho = \rho -1 = {\cal O}(10^{-3})$~\cite{PDG}. 
Assuming a small Majorana mass term, $y v_s/\sqrt{2}$, 
 we evaluate 1-loop corrections to $\Delta \rho$ 
 via the dark matter fermions with the mass matrix 
 in Eq~(\ref{M-Dirac}). 
For $\mu_\chi > Y v/\sqrt{2} $, we find 
\begin{eqnarray}
\Delta \rho \simeq 
 \frac{Y^2}{4 \pi^2} \left( \frac{\mu_\psi}{\mu_\chi-\mu_\psi}\right). 
\end{eqnarray}
The constraint is satisfied for $Y \lesssim 0.3$ 
 with $\mu_\chi \gtrsim 3 \mu_\psi$. 
Thus, our discussion on the direct/indirect detection of dark matter 
 is consistent  with the $\rho$ parameter constraint.

\section{Model B}

\subsection{Setup}

Next we consider another realization of our scenario 
 in the context of a gauged $U(1)_{B-L}$ extended model.  
In addition to the minimal gauged $U(1)_{B-L}$ extension 
 with the right-handed neutrinos $N_R$ and $B-L$ Higgs 
 field $\Psi$~\cite{Emam:2007dy,Huitu:2008gf,Khalil:2008kp,Basso:2008iv,Iso:2009ss,Perez:2009mu},
 we further introduce a vectorlike pair of SM singlet fermions 
 ($\psi_{L,R}$)
 with $B-L$ charge $-2$, another vector-like pair of 
 totally singlet fermions ($\chi_{L,R}$) 
 and two $B-L$ charged/uncharged complex scalars ($\eta$ and $\phi$). 
The charge assignments are shown in Table~II. 
Here we have introduced a discrete $Z_2$ parity to ensure 
 the stability of dark matter particle and 
 a global $U(1)$ symmetry to forbid Majorana mass terms   
 for the $B-L$ charge neutral fermions.

\begin{table}[ht]
\caption{Particle contents of Model B}
\centering
\begin{center}
\begin{tabular}{|l|c|c|c|c|c|} \hline
Fields  & $SU(2)_L$ & $U(1)_Y$ & $U(1)_{B-L}$ & $Z_{2}$ parity & Global U(1)   \\ \hline\hline
$\psi_L$ &   1       &     0       &     $-2$        &    $ - $          &    $-1$          \\ \hline
$\psi_R$ &   1       &     0       &     $-2$        &    $ - $          &    $-1$          \\ \hline
$\chi_L$ &   1       &     0       &      $0$         &    $ - $          &    $-1$          \\ \hline
$\chi_R$ &   1       &     0       &      $0$         &    $ - $          &    $-1$          \\ \hline\hline
$\Psi$    &   1       &    0        &      $-2$       &    $ + $          &    0               \\ \hline
$\eta$    &   1       &    0        &      $+1$       &    $ - $          &    $+1$          \\ \hline
$\phi$    &   1       &    0        &      $0$         &    $ + $          &    $+2$          \\ \hline\hline
$N_R$     &   1       &    0        &      $-1$       &    $ + $          &    0               \\ \hline
\end{tabular}
\end{center}
\end{table}

The Lagrangian relevant for our discussion is given by
\begin{eqnarray}
 -{\cal L} & \supset &  \frac{1}{2}y \bar{\chi_R^c} \phi \chi_R 
+ \frac{1}{2}y \bar{\chi_L^c} \phi \chi_L
 + \frac{1}{2} \bar{N_R^c} (Y_{\Psi} \Psi^{\dagger}) N_R \nonumber \\
  & & + \mu_{\chi} \bar{\chi_L} \chi_R + \mu_{\psi} \bar{\psi_L} \psi_R 
 + Y \bar{\psi_L} \Psi \chi_R + Y \bar{\chi_L} \Psi^\dagger \psi_R 
  + Y_N \bar{\psi_L} \eta^{\dagger} N_R
 + {\it h.c.} \nonumber  \\
  && + V(\Phi, \Psi, \eta, \phi)
\end{eqnarray}
 with
\begin{eqnarray}
V(\Phi, \Psi, \eta, \phi) & = & m_1^2 |\Phi|^2 + m_2^2 |\Psi|^2 +\lambda_1 |\Phi|^4 
  + \lambda_2|\Psi|^4 + \lambda_3 |\Phi|^2|\Psi|^2  \nonumber \\
  && + m_3^2 |\phi|^2 + (\lambda_4 |\Phi|^2 + \lambda_5|\Psi|^2+ \lambda_6 |\eta|^2 ) |\phi|^2 
       + \lambda_7|\phi|^4 \nonumber \\
  && + m_4^2 |\eta|^2  +  ( \lambda_8 |\Phi|^2  + \lambda_9 |\Psi|^2 ) |\eta|^2
       + \lambda_{10}|\eta|^4  \nonumber \\
  && + \lambda_W \eta\eta \Psi\phi^{\dagger } + {\it h.c.} ,
\end{eqnarray}
 where $\Phi$ denotes the SM Higgs doublet field. 
For simplicity, we have assumed the universal Yukawa coupling
 constants $y$ and $Y$.  
We fix the parameters in the scalar potential 
 so as for $\Phi, \Psi$ and $\phi$ to develop VEVs 
 and in this case the physical Higgs bosons are given by 
\begin{eqnarray}
 \Phi = \left( 
            \begin{array}{c}
            0 \\
            \frac{1}{\sqrt{2}}(v + h) 
            \end{array}
              \right) , \; 
 \Psi = \frac{1}{\sqrt{2}}(v' + H) , \; 
 \phi = \frac{1}{\sqrt{2}}(v_s + \varphi) 
\end{eqnarray}
 around their VEVs ($v$, $v'$ and $v_s$). 
The mass eigenstates are obtained by the unitary rotation 
\begin{eqnarray}
 \left( 
            \begin{array}{c}
            H_1 \\
            H_2 \\
            H_3
            \end{array}
 \right)
   &=&
 \left( 
            \begin{array}{ccc}
            S_{11} & S_{12} & S_{13}  \\
            S_{21} & S_{22} & S_{23}  \\
            S_{31} & S_{32} & S_{33}  
            \end{array}
 \right) 
   \left( 
            \begin{array}{c}
            \varphi \\
            h \\
            H 
            \end{array}
  \right) ,
\end{eqnarray}
 by which the mass matrix 
\begin{equation}
 \textrm{ Mass term } = 
  \frac{1}{2} \left( \varphi \,\,  h \,\,  H  \right)
  \left(
\begin{array}{ccc}
 2 v_s^2 \lambda_7 & v v_s \lambda_4 & v' v_s \lambda_5 \\
 v v_s \lambda_4 &  2 v^2 \lambda_1  & v v' \lambda_3  \\
 v' v_s \lambda_5 & v v' \lambda_3    & 2 v'^2 \lambda_2
\end{array}
\right)
   \left( 
            \begin{array}{c}
            \varphi \\
            h \\
            H 
            \end{array}
  \right) 
\end{equation}
 is diagonalized.

\subsection{Cosmological evolution}

When $\Psi$ develops the VEV $\langle\Psi\rangle = \frac{v'}{\sqrt{2}}$,
 the $B-L$ gauge symmetry is broken. 
Associated  with the symmetry breaking, the $B-L$ gauge boson ($Z'$) 
 and the right-handed neutrinos acquire their masses, 
 $m_{Z'}^2 = 4 g_{B-L}^2 v'^2 $ and 
 $m_{N_R} = \frac{Y_{\Psi} v'}{\sqrt{2}}$. 
We set this symmetry breaking at a energy much higher 
 than the electroweak symmetry breaking. 
The current lower bound on $v'$ is found to be 
 $v' \gtrsim 3$ TeV~\cite{Carena:2004xs, Cacciapaglia:2006pk}, 
 and our setup is consistent with it.

Right-handed neutrinos have two decay modes:
 decays into lepton and Higgs ($N_R \rightarrow L \Phi $) 
 and into $\eta$ and $\psi$ ($N_R \rightarrow \psi \eta $).
As the same as the discussion for Model A in the previous section, 
 the CP-asymmetric decays of the right-handed neutrinos 
 generate the lepton asymmetry and the asymmetry of $\psi$ and $\eta$. 
We assume the scalar $\eta$ is the heavier than 
 the other $Z_2$-odd particles. 
The $\psi$ asymmetry generated by $N_R$ decay 
 is nothing but the asymmetry of the dark matter sector. 
Notice that the same amount of asymmetry 
 between $\eta$ and $\eta^*$ is also generated as in Model A.

After the $B-L$ symmetry is broken by $\langle \Phi \rangle=v'/\sqrt{2}$
 (but $\phi$ has not developed the VEV at this moment), 
 $Z_2$-odd Dirac fermion mass matrix is given by 
\begin{equation}
 (\bar{\chi_L}  \bar{\chi_R^c}  \bar{\psi_L}  \bar{\eta_R^c})
 M
\left(
\begin{array}{c}
 \chi_L^c \\
 \chi_R \\
 \psi_L^c  \\
 \psi_R \\
\end{array}
\right)
\end{equation}
 with 
\begin{eqnarray}
 M=
\left(
\begin{array}{cccc}
 y \langle \phi \rangle  & \mu_{\chi}  & 0 & Y \frac{v'}{\sqrt{2}}  \\
 \mu_{\chi} & y \langle \phi \rangle & Y \frac{v'}{\sqrt{2}} & 0   \\
 0 & Y \frac{v'}{\sqrt{2}} & 0 & \mu_{\psi}   \\
 Y \frac{v'}{\sqrt{2}} & 0 & \mu_{\psi} & 0   \\
\end{array}
\right), 
\label{Mf}
\end{eqnarray}
where $ \langle \phi \rangle =0$. 
The mass eigenstates are mixture of two Dirac fermions, 
 $\chi$ and $\psi$ and their mass eigenvalues are given by 
\begin{eqnarray} 
 M^{Dirac}_{\chi\psi} = \frac{\mu_{\chi}+\mu_{\psi} \pm \sqrt{ 2 Y^2 v'^2 + ( \mu_{\chi} - \mu_{\psi})^2 } }{2} .
\end{eqnarray}
The lighter mass eigenstate is nothing but the asymmetric 
 Dirac fermion dark matter.

The WIMP dark matter scenario in the context of the gauged $B-L$ model 
 has been investigated before, and it has been shown that 
 the dark matter annihilation processes are not efficient.  
Only in the limited cases with 
 the Higgs boson resonance~\cite{Okada:2010wd,Kanemura:2011vm,Okada:2012sg} 
 or the the $Z'$ boson resonance~\cite{Burell:2011wh,Okada:2012sg}, 
 the relic abundance of the dark matter particle can be lower than 
 the observed value. 
Therefore, in order to realize our asymmetric dark matter scenario, 
 we need to tune the dark matter masses $M^{Dirac}_{\chi\psi}$ 
 to be close to half of either Higgs bosons or $Z'$ boson masses,
 and so is the $\eta$ mass in order to realize 
 efficient annihilations between $\eta$ and $\eta^*$.

At a low energy where only $\phi$ is light and the others are
decoupled,  
 the scalar potential is simplified as 
\begin{eqnarray}
V(\Phi, \eta, \phi) \simeq  
  - m_\phi^2 |\phi|^2 + \lambda_7|\phi|^4,  
\end{eqnarray} 
 where $m_\phi^2 = - m_3^2 - \frac{\lambda_4 v^2 + \lambda_5 v'^2}{2} > 0$, 
 and the scalar $\phi$ develops the VEV. 
In the early Universe, we consider the thermal effects on 
 the scalar potential. 
By the same analysis in the previous section, 
 the thermal  mass term for $\phi$ is found to be 
 $m_{th}^2 \simeq T^2 \lambda_7/6$ in the high-temperature expansions.
Thus, the phase transition occurs at the temperature 
 $T_c \simeq \sqrt{6 m_\phi^2/\lambda_7}$. 
Again, we assume a small $m_\phi$ and this phase 
 transition takes place well after the dark matter 
 and antidark matter annihilation have frozen-out. 
The same arguments on cosmic strings and NG mode for Model A 
 are applicable to this model.

As the same as in the Model A, 
 the $\eta$ decay produce antidark matter at late time. 
Since this happens after the freeze out of 
 the annihilation processes of dark matters and $\eta$s, 
 the produced antidark matters remain without annihilations 
 with the same number density as the dark matter one.

\subsection{Direct dark matter detection}
Once the $\phi$ has developed the VEV in a late time, 
 the fermion mass matrix in Eq.~(\ref{Mf}) has nonzero 
 diagonal elements and the dark matter particle 
 has  taken the form of the Majorana fermion. 
Because of the Majorana nature, the dark matter particle 
 has the axial vector coupling with the $Z'$ boson, 
 while the SM quarks have the vector coupling. 
As a result, there is no effective couplings between 
 the dark matter and the SM quarks induced by the $Z'$ boson exchange 
 in the nonrelativistic limit. 
Therefore, for Model B, we only consider 
 the spin-independent elastic scattering process 
 via Higgs boson exchanges.

The coefficient of the operator ${\cal L}_{int} 
 = \alpha_q \chi_j \chi_j \bar{q} q $ is found to be 
\begin{eqnarray}
\alpha_q = \frac{1}{2}\sum_i \frac{ (N_{j1}^2+N_{j2}^2) y S_{i1} + 2(N_{j2}N_{j3}+ N_{j1}N_{j4}) Y S_{i3}}{\sqrt{2} v M_{H_i}^2 }m_q S_{i2} ,
\end{eqnarray}
 where $N_{ij}$ is the unitary matrix to diagonalize the mass matrix. 
In the limit $y v_s \ll Y v'$ for simplicity, 
 $\alpha_q$ is reduced as
\begin{eqnarray}
\alpha_q \simeq \frac{1}{2}\sum_i \frac{ y S_{i1} \pm 2 Y S_{i3}}{2 \sqrt{2} v M_{H_i}^2 }m_q S_{i2} .
\end{eqnarray}

\subsection{Benchmark scenarios}
As we have discussed above, viable parameter sets are very limited 
 in order to achieve the large annihilation cross section  
 for dark matter particles via the enhancement 
 of the Higgs or $Z'$ boson resonances. 
Let us consider two benchmarks.  

\subsubsection{The $Z'$ boson resonance}
\label{B:ZprimeResonance}
We take the asymmetric dark matter mass to be 
 $M^{Dirac}_{\chi \psi} \simeq M_{Z'}/2$, 
 so that the dark matter annihilation cross section 
 is enhanced~\cite{Burell:2011wh, Okada:2012sg} 
 and the relic dark matter originates 
 from the dark matter asymmetry in the Universe. 
Similarly, we take the $\eta$ mass to be 
 around half of the heavy Higgs boson mass. 
In this case there is little correlation between
 the annihilation cross section and the spin-independent 
 cross section of the dark matter off the nuclei, 
 except the dark matter mass being half of $Z'$ boson mass.

\subsubsection{A Higgs boson resonance}
\label{B:HiggsResonance}
Another benchmark is to fix the dark matter mass 
 $M^{Dirac}_{\chi \psi} \simeq M_{H_i}/2$ 
 and $M_{\eta} \simeq M_{H_j}/2$ or $M_{Z'}/2$, 
 where the Dirac fermion annihilation cross section is enhanced 
 by the $s$-channel Higgs resonances and 
 $\eta$ annihilation cross section is enhanced 
 by the $s$-channel Higgs or $Z'$ resonances 
 as discussed in Refs.~\cite{Okada:2010wd,Kanemura:2011vm} 
 for similar models. 
For simplicity, we fix the dark matter mass to be 
 half of the SM-like Higgs boson and in this case 
 the structure of our model is similar to 
 the so-called Higgs portal dark matter scenario. 
A representative model is the SM plus a gauge singlet real 
 scalar as the dark matter~\cite{
 McDonald:1993ex,Burgess:2000yq,Davoudiasl:2004be,Kikuchi:2007az}. 
Since the sensitivity of the dark matter direct detection 
 experiments has been greatly improved, 
 the allowed parameter region of the model to simultaneously 
 satisfy the constraints from the relic abundance and the direct
 detection is very limited~\cite{He:2009yd,Aoki:2009pf,Kanemura:2010sh,Djouadi:2011aa}: 
For relatively light dark matter, say, $M_{DM} \lesssim 1$ TeV, 
 the dark matter mass should be around half of 
 the Higgs boson mass ($M_{DM} \approx M_H/2)$. 
The correct relic abundance is achieved by the Higgs resonance 
 even though the coupling between the dark matter and 
 the Higgs boson is very small. 
On the other hand, such a small coupling makes the direct detection 
 of the dark matter very hard. 
If the dark matter is lighter than half of the Higgs mass 
 and the dark matter mass is not just on the Higgs pole, 
 we can find the Higgs portal dark matter signal 
 at high energy colliders such 
 as the LHC~\cite{Kanemura:2010sh,Djouadi:2011aa} 
 through the invisible decay of the Higgs boson 
 into the dark matter particle pair~\cite{Bento:2000ah,Bento:2001yk}.

In our model, there is a crucial difference from 
 the Higgs portal dark matter. 
For the asymmetric dark matter scenario, the large annihilation 
 cross section is welcome in order to completely erase 
 the symmetric part of the relic abundance of dark matter 
 and antidark matter, leaving only the asymmetric part. 
Therefore, it is not necessary for the coupling 
 between the dark matter and the Higgs boson to be small. 
In this case the spin-independent cross section 
 can be accessible to the future experiments.

\section{Summary and Discussions}

We have proposed a novel asymmetric dark matter scenario. 
The dark matter particle is initially the Dirac fermion 
 and the dark matter asymmetry is generated 
 by the same mechanism as leptogenesis. 
Through efficient dark matter annihilation processes, 
 the symmetric part of the dark matter number density 
 is erased and the dark matter relic density is 
 determined by the dark matter asymmetry. 
After the annihilation processes have frozen out, 
 a phase transition of a light scalar field takes place 
 and generates Majorana mass terms for the dark matter particles. 
As a result, the originally asymmetric dark matter 
 turns into the Majorana fermion. 
Although the dark matter behaves as the WIMP at present, 
 its relic abundance is basically independent of 
 the pair annihilation cross section.

In order to address this scenario in detail, 
 we have proposed two simple models. 
The first model is based on the SM gauge group 
 and no new gauge interaction is introduced. 
The dark matter originates from the $SU(2)_L$ doublet  
 Dirac fermion and similar to the Higgsino in the MSSM. 
Through the coupling with the right-handed neutrinos, 
 the dark matter asymmetry in the Universe is 
 generated by the out-of-equilibrium decay 
 of the right-handed neutrinos, the same mechanism 
 as leptogenesis. 
The annihilation cross section through the $SU(2)_L$ 
 gauge interaction is large enough to erase 
 the symmetric part of the dark matter and antidark matter
 abundance. 
At a low energy after the freeze-out of the annihilation process, 
 the thermal effects for the scalar potential becomes smaller, 
 and a light scalar field develops the VEV. 
The Majorana mass term generated by the VEV turns 
 the dark matter particle into the Majorana particle. 
Due to this Majorana nature, the dark matter has 
 no spin-independent interaction with the quarks 
 mediated by the $Z$ boson. 
We have estimated the spin independent and independent  
 elastic scattering cross sections of the dark matter 
 with a proton and found that the cross sections 
 can be close to the current experimental bounds 
 with reasonable model-parameter choices. 
Therefore, the dark matter can be detected in the near future.

In the second model, we have introduced an extra $U(1)_{B-L}$ gauge 
 interaction and all new particles introduced are singlet 
 under the SM gauge group. 
An efficient Dirac fermion dark matter annihilation 
 is achieved through the resonance effect by 
 the Higgs or $Z'$ boson mediated processes. 
As a result, the dark matter mass should be around
 half of either Higgs or $Z'$ boson masses. 
As in the first model, the phase transition of a light scalar 
 occurs after the annihilation processes have been frozen-out, 
 and the dark matter takes the form of the Majorana fermion 
 at a late time. 
This second model is similar to the Higgs portal dark matter scenario. 
A crucial difference is that there is no constraint 
 on the coupling between the dark matter and Higgs bosons 
 from the dark matter relic abundance, 
 because the relic abundance is determined 
 by the dark matter asymmetry. 
Therefore, the spin-independent cross section can be 
 as large as the current upper limit from the direct dark matter 
 search experiments, independently of the annihilation cross section. 
This situation is similar to the nonthermal dark matter scenario, 
 where the dark matter particle is produced by 
 unstable relics such as moduli~\cite{Moroi:1999zb},
 Q-ball~\cite{Fujii:2001xp}, and axino~\cite{Choi:2008zq}.

Finally, we note that the realization of our scenario 
 in the context of MSSM is challenging 
 because the Majorana mass terms for the gauginos is always present. 
Recently, it has been shown~\cite{Blum:2012nf} 
 that an extreme parameter choice can realize 
 the Higgsino to be a viable asymmetric dark matter candidate.

%
%
\section*{Acknowledgments}
O.S. would like to thank the Department of Physics and Astronomy, 
 University of Alabama for their warm hospitality during his visit. 
This work is in part supported by the DOE Grants 
 No. DE-FG02-10ER41714 (N.O.) and by the scientific 
 research grants from Hokkai-Gakuen University (O.S.).

\appendix

\section{} 

Here, we note the explicit form of $N$.
\begin{eqnarray}
 N =
\left(
 \begin{array}{cccc}
    -\frac{Y v}{ \sqrt{ (2 Y v)^2 + ( y v_s )^2 }D_+} & \frac{Y v}{ \sqrt{ (2 Y v)^2 + ( y v_s )^2 }D_+}  & -\frac{1}{2} D_+ & \frac{1}{2} D_+  \\
    \frac{Y v}{ \sqrt{ (2 Y v)^2 + ( y v_s )^2 }D_-} & -\frac{Y v}{ \sqrt{ (2 Y v)^2 + ( y v_s )^2 }D_-}  & -\frac{1}{2} D_- & \frac{1}{2} D_-  \\ 
    -\frac{Y v}{ \sqrt{ (2 Y v)^2 + ( y v_s )^2 }D_+} & -\frac{Y v}{ \sqrt{ (2 Y v)^2 + ( y v_s )^2 }D_+}  & \frac{1}{2} D_+ & \frac{1}{2} D_+  \\
    \frac{Y v}{ \sqrt{ (2 Y v)^2 + ( y v_s )^2 }D_-} & \frac{Y v}{ \sqrt{ (2 Y v)^2 + ( y v_s )^2 }D_-}  & \frac{1}{2} D_- & \frac{1}{2} D_-
 \end{array}
\right) ,
\end{eqnarray}
 with
\begin{eqnarray}
D_+ &=& \sqrt{1 + \frac{y v_s}{\sqrt{ (2 Y v)^2 + ( y v_s )^2 }}} , \\
D_- &=& \sqrt{1 - \frac{y v_s}{\sqrt{ (2 Y v)^2 + ( y v_s )^2 }}} .
\end{eqnarray}
%

\section{Amplitude}

%

We give explicit formulas of the invariant amplitude 
 squared for the pair annihilation of Dirac dark matter $\chi$
 into light singlet scalar $\phi$ pair. 
\begin{eqnarray}
&&  w(s; \rightarrow \phi\phi)
 \equiv \frac{1}{4}\int d{\rm LIPS} \overline{|{\cal M}|}^2
 \simeq \frac{1}{32 \pi}  
 \left(N y N \right)^4
  \left(
   F(s) + G(s) \ln\left|\frac{a+b}{a- b}\right|
  \right) , 
\end{eqnarray}
The auxiliary functions that appear above are defined as
\begin{eqnarray} 
F(s)
 &\equiv& 2 \left(-2 - \frac{(4 m_{\chi}^2- m_{\phi}^2)^2}{m_{\chi}^2 (s -4 m_{\phi}^2) + m_{\phi}^4} \right)  , \\
G(s)
 &\equiv&  2 \sqrt{\frac{ s-4 m_{\phi}^2 }{ s-4 m_{\chi}^2 } }
 \frac{ \left( -32m_{\chi}^4-4 m_{\phi}^2(s+4 m_{\chi}^2) + 16 s m_{\chi}^2 + 6 m_{\phi}^4 + s^2 \right)}
 {s^2- 6 s m_{\phi}^2 + 8 m_{\phi}^4} , \\
 a(s) &\equiv& s - 2 m_{\phi}^2 , \\
 b(s) &\equiv& \sqrt{ (s- 4 m_{\phi}^2)(s- 4 m_{\chi}^2)} .
\end{eqnarray}
%




\begin{thebibliography}{99}

\bibitem{Komatsu:2010fb}
  E.~Komatsu {\it et al.}  [WMAP Collaboration],
  Astrophys.\ J.\ Suppl.\  {\bf 192}, 18 (2011).


\bibitem{Enqvist:1997si}
  K.~Enqvist and J.~McDonald,
  Phys.\ Lett.\  B {\bf 425}, 309 (1998).
  
\bibitem{Enqvist:1998en}
  K.~Enqvist and J.~McDonald,
  Nucl.\ Phys.\  B {\bf 538}, 321 (1999).
  
\bibitem{Enqvist:2000gq}
  K.~Enqvist, A.~Jokinen and J.~McDonald,
  Phys.\ Lett.\  B {\bf 483}, 191 (2000).
  
\bibitem{Roszkowski:2006kw}
  L.~Roszkowski and O.~Seto,
  Phys.\ Rev.\ Lett.\  {\bf 98}, 161304 (2007).

\bibitem{Seto:2007ym}
  O.~Seto and M.~Yamaguchi,
  Phys.\ Rev.\  D {\bf 75}, 123506 (2007).

\bibitem{Shoemaker:2009kg}
  I.~M.~Shoemaker and A.~Kusenko,
  Phys.\ Rev.\ D {\bf 80}, 075021 (2009).







\bibitem{Barr:1990ca}
  S.~M.~Barr, R.~S.~Chivukula and E.~Farhi,
  Phys.\ Lett.\  B {\bf 241}, 387 (1990).

\bibitem{Barr:1991qn}
  S.~M.~Barr,
  Phys.\ Rev.\  D {\bf 44}, 3062 (1991).

\bibitem{Kaplan:1991ah}
  D.~B.~Kaplan,
  Phys.\ Rev.\ Lett.\  {\bf 68}, 741 (1992).

\bibitem{Thomas:1995ze}
  S.~D.~Thomas,
  Phys.\ Lett.\  B {\bf 356}, 256 (1995).

\bibitem{Hooper:2004dc}
  D.~Hooper, J.~March-Russell and S.~M.~West,
  Phys.\ Lett.\  B {\bf 605}, 228 (2005).

\bibitem{Kitano:2004sv}
  R.~Kitano and I.~Low,
  Phys.\ Rev.\  D {\bf 71}, 023510 (2005).

\bibitem{Agashe:2004bm}
  K.~Agashe and G.~Servant,
  JCAP {\bf 0502}, 002 (2005).

\bibitem{Cosme:2005sb}
  N.~Cosme, L.~Lopez Honorez and M.~H.~G.~Tytgat,
  Phys.\ Rev.\  D {\bf 72}, 043505 (2005).

\bibitem{Farrar:2005zd}
  G.~R.~Farrar and G.~Zaharijas,
  Phys.\ Rev.\ Lett.\  {\bf 96}, 041302 (2006).

\bibitem{Khlopov:2005ew}
  M.~Y.~Khlopov,
  Pisma Zh.\ Eksp.\ Teor.\ Fiz.\  {\bf 83}, 3 (2006)
  [JETP Lett.\  {\bf 83}, 1 (2006)].

\bibitem{Kaplan:2009ag}
  D.~E.~Kaplan, M.~A.~Luty and K.~M.~Zurek,
  Phys.\ Rev.\  D {\bf 79}, 115016 (2009).

\bibitem{Cohen:2010kn}
  T.~Cohen, D.~J.~Phalen, A.~Pierce and K.~M.~Zurek,
  Phys.\ Rev.\  D {\bf 82}, 056001 (2010).
\bibitem{Buckley:2011kk}
  M.~R.~Buckley,
  Phys.\ Rev.\  D {\bf 84}, 043510 (2011).
\bibitem{Graesser:2011wi}
  M.~L.~Graesser, I.~M.~Shoemaker and L.~Vecchi,
  JHEP {\bf 1110}, 110 (2011).
\bibitem{Iminniyaz:2011yp}
  H.~Iminniyaz, M.~Drees and X.~Chen,
  JCAP {\bf 1107}, 003 (2011).
\bibitem{MarchRussell:2012hi} 
  J.~March-Russell, J.~Unwin and S.~M.~West,
  arXiv:1203.4854 [hep-ph].

\bibitem{Ibe:2011hq}
  M.~Ibe, S.~Matsumoto and T.~T.~Yanagida,
  Phys.\ Lett.\ B {\bf 708}, 112 (2012).

\bibitem{HM}
  M.~Beck {\it et al.},
  Phys.\ Lett.\  B {\bf 336}, 141 (1994).

\bibitem{Cohen:2009fz}
  T.~Cohen and K.~M.~Zurek,
  Phys.\ Rev.\ Lett.\  {\bf 104}, 101301 (2010).
\bibitem{Buckley:2011ye}
  M.~R.~Buckley and S.~Profumo,
  Phys.\ Rev.\ Lett.\  {\bf 108}, 011301 (2012).
\bibitem{Cirelli:2011ac}
  M.~Cirelli, P.~Panci, G.~Servant and G.~Zaharijas,
  JCAP {\bf 1203}, 015 (2012). 
\bibitem{Tulin:2012re}
  S.~Tulin, H.~B.~Yu and K.~M.~Zurek,
  arXiv:1202.0283 [hep-ph].

\bibitem{An:2009vq}
  H.~An, S.~L.~Chen, R.~N.~Mohapatra and Y.~Zhang,
  JHEP {\bf 1003}, 124 (2010).
\bibitem{Haba:2010bm}
  N.~Haba and S.~Matsumoto,
  Prog.\ Theor.\ Phys.\  {\bf 125}, 1311 (2011).
\bibitem{Blennow:2010qp}
  M.~Blennow, B.~Dasgupta, E.~Fernandez-Martinez and N.~Rius,
  JHEP {\bf 1103}, 014 (2011).
\bibitem{Falkowski:2011xh}
  A.~Falkowski, J.~T.~Ruderman and T.~Volansky,
  JHEP {\bf 1105}, 106 (2011).


\bibitem{Bell:2011tn}
  N.~F.~Bell, K.~Petraki, I.~M.~Shoemaker and R.~R.~Volkas,
  Phys.\ Rev.\  D {\bf 84}, 123505 (2011).
\bibitem{Cheung:2011if}
  C.~Cheung and K.~M.~Zurek,
  Phys.\ Rev.\  D {\bf 84}, 035007 (2011).
\bibitem{Graesser:2011vj}
  M.~L.~Graesser, I.~M.~Shoemaker and L.~Vecchi,
  arXiv:1107.2666 [hep-ph].
\bibitem{vonHarling:2012yn}
  B.~von Harling, K.~Petraki and R.~R.~Volkas,
  arXiv:1201.2200 [hep-ph].

\bibitem{MarchRussell:2011fi}
  J.~March-Russell and M.~McCullough,
  JCAP {\bf 1203}, 019 (2012). 
\bibitem{Kamada:2012ht}
  K.~Kamada and M.~Yamaguchi,
  arXiv:1201.2636 [hep-ph].


\bibitem{FukugitaYanagida}
M.~Fukugita and T.~Yanagida,
 Phys.\ Lett.\ B {\bf 174}, 45 (1986).

\bibitem{ThermalQFT}
  e.g., J.~I.~Kapusta, {\it Finite-temperature field theory},
  Cambridge University Press (1994); \\
  M.~Le Bellac, {\it Thermal Field Theory},
  Cambridge University Press (2000).
  
\bibitem{seesaw}
P.~Minkowski, Phys. Lett. B {\bf 67}, 421 (1977);
T.~Yanagida, in \emph{Proceedings of the Workshop on the Unified
  Theory and the Baryon Number in the Universe} (O.~Sawada and
  A.~Sugamoto, eds.), KEK, Tsukuba, Japan, 1979, p.~95;
M.~Gell-Mann, P.~Ramond, and R.~Slansky, \emph{Supergravity} (P.~van
  Nieuwenhuizen et al. eds.), North Holland, Amsterdam, 1979, p.~315;
S.~L. Glashow, \emph{The future of elementary particle physics}, in
  \emph{Proceedings of the 1979 Carg{\`e}se Summer Institute
 on Quarks and Leptons} (M.~Levy et al. eds.),
 Plenum Press, New York, 1980, p.~687;
R.~N. Mohapatra and G.~Senjanovic,
 Phys. Rev. Lett. {\bf 44}, 912 (1980).




\bibitem{Keisler:2011aw} 
  R.~Keisler, C.~L.~Reichardt, K.~A.~Aird, B.~A.~Benson, L.~E.~Bleem, J.~E.~Carlstrom, C.~L.~Chang and H.~M.~Cho {\it et al.},
  Astrophys.\ J.\  {\bf 743}, 28 (2011).


\bibitem{Falk:1994es}
  T.~Falk, K.~A.~Olive and M.~Srednicki,
  Phys.\ Lett.\  B {\bf 339}, 248 (1994).

\bibitem{ArkaniHamed:2000bq}
  N.~Arkani-Hamed, L.~J.~Hall, H.~Murayama, D.~Tucker-Smith and N.~Weiner,
  Phys.\ Rev.\  D {\bf 64}, 115011 (2001).

\bibitem{Lee:2007mt}
  H.~S.~Lee, K.~T.~Matchev and S.~Nasri,
  Phys.\ Rev.\  D {\bf 76}, 041302 (2007).
\bibitem{Cerdeno:2008ep}
  D.~G.~Cerdeno, C.~Munoz and O.~Seto,
  Phys.\ Rev.\  D {\bf 79}, 023510 (2009).


\bibitem{Aprile:2011hi}
  E.~Aprile {\it et al.}  [XENON100 Collaboration],
  Phys.\ Rev.\ Lett.\  {\bf 107}, 131302 (2011).

\bibitem{Girard:2011xc}
  T.~Girard, T.~A.~Morlat, M.~Felizardo, A.~C.~Fernandes, A.~R.~Ramos and J.~G.~Marques
                  [for the SIMPLE Collaboration],
  PoS {\bf IDM2010}, 055 (2011).

\bibitem{Abbasi:2011zzr}
  R.~Abbasi {\it et al.}  [IceCube Collaboration],
  arXiv:1111.2738 [astro-ph.HE].

\bibitem{PDG}
K.~Nakamura {\it et al.}  [Particle Data Group Collaboration],
 J.\ Phys.\ G G {\bf 37}, 075021 (2010).


  

\bibitem{Emam:2007dy}
  W.~Emam and S.~Khalil,
  Eur.\ Phys.\ J.\  C {\bf 522}, 625 (2007).

\bibitem{Huitu:2008gf}
  K.~Huitu, S.~Khalil, H.~Okada and S.~K.~Rai,
  Phys.\ Rev.\ Lett.\  {\bf 101}, 181802 (2008).

\bibitem{Khalil:2008kp}
  S.~Khalil and O.~Seto,
  JCAP {\bf 0810}, 024 (2008).

\bibitem{Basso:2008iv}
  L.~Basso, A.~Belyaev, S.~Moretti and C.~H.~Shepherd-Themistocleous,
  Phys.\ Rev.\  D {\bf 80}, 055030 (2009).

\bibitem{Iso:2009ss}
  S.~Iso, N.~Okada and Y.~Orikasa,
  Phys.\ Lett.\  B {\bf 676}, 81 (2009); 
%
  Phys.\ Rev.\  D {\bf 80}, 115007 (2009). 

\bibitem{Perez:2009mu}
  P.~F.~Perez, T.~Han and T.~Li,
  Phys.\ Rev.\  D {\bf 80}, 073015 (2009).

\bibitem{Carena:2004xs} 
  M.~S.~Carena, A.~Daleo, B.~A.~Dobrescu and T.~M.~P.~Tait,
  Phys.\ Rev.\ D {\bf 70}, 093009 (2004).

\bibitem{Cacciapaglia:2006pk} 
  G.~Cacciapaglia, C.~Csaki, G.~Marandella and A.~Strumia,
  Phys.\ Rev.\ D {\bf 74}, 033011 (2006).



\bibitem{Okada:2010wd}
  N.~Okada and O.~Seto,
  Phys.\ Rev.\  D {\bf 82}, 023507 (2010).

\bibitem{Kanemura:2011vm}
  S.~Kanemura, O.~Seto and T.~Shimomura,
  Phys.\ Rev.\  D {\bf 84}, 016004 (2011).

\bibitem{Okada:2012sg}
  N.~Okada and Y.~Orikasa,
  arXiv:1202.1405 [hep-ph].
  
\bibitem{Burell:2011wh}
  Z.~M.~Burell and N.~Okada,
  Phys.\ Rev.\ D {\bf 85}, 055011 (2012). 

\bibitem{McDonald:1993ex}
  J.~McDonald,
  Phys.\ Rev.\  D {\bf 50}, 3637 (1994).

\bibitem{Burgess:2000yq}
  C.~P.~Burgess, M.~Pospelov and T.~ter Veldhuis,
  Nucl.\ Phys.\  B {\bf 619}, 709 (2001).

\bibitem{Davoudiasl:2004be}
  H.~Davoudiasl, R.~Kitano, T.~Li and H.~Murayama,
  Phys.\ Lett.\  B {\bf 609}, 117 (2005).

\bibitem{Kikuchi:2007az}
  T.~Kikuchi and N.~Okada,
  Phys.\ Lett.\  B {\bf 665}, 186 (2008). 

%

\bibitem{He:2009yd}
  X.~G.~He, T.~Li, X.~Q.~Li, J.~Tandean and H.~C.~Tsai,
  Phys.\ Lett.\  B {\bf 688}, 332 (2010).

\bibitem{Aoki:2009pf}
  M.~Aoki, S.~Kanemura and O.~Seto,
  Phys.\ Lett.\  B {\bf 685}, 313 (2010).

\bibitem{Kanemura:2010sh}
  S.~Kanemura, S.~Matsumoto, T.~Nabeshima and N.~Okada,
  Phys.\ Rev.\  D {\bf 82}, 055026 (2010).

\bibitem{Djouadi:2011aa}
  A.~Djouadi, O.~Lebedev, Y.~Mambrini and J.~Quevillon,
  Phys.\ Lett.\ B {\bf 709}, 65 (2012). 

\bibitem{Bento:2000ah}
  M.~C.~Bento, O.~Bertolami, R.~Rosenfeld and L.~Teodoro,
  Phys.\ Rev.\  D {\bf 62}, 041302 (2000).

\bibitem{Bento:2001yk}
  M.~C.~Bento, O.~Bertolami and R.~Rosenfeld,
  Phys.\ Lett.\  B {\bf 518}, 276 (2001).



\bibitem{Moroi:1999zb}
  T.~Moroi and L.~Randall,
  Nucl.\ Phys.\  B {\bf 570}, 455 (2000).

\bibitem{Fujii:2001xp}
  M.~Fujii and K.~Hamaguchi,
  Phys.\ Lett.\  B {\bf 525}, 143 (2002).

\bibitem{Choi:2008zq}
  K.~Y.~Choi, J.~E.~Kim, H.~M.~Lee and O.~Seto,
  Phys.\ Rev.\  D {\bf 77}, 123501 (2008).

\bibitem{Blum:2012nf}
  K.~Blum, A.~Efrati, Y.~Grossman, Y.~Nir and A.~Riotto,
  arXiv:1201.2699 [hep-ph].



\end{thebibliography}
\end{document}